\documentclass[3p]{elsarticle}

\usepackage{hyperref}
\usepackage{color}
\usepackage{graphicx}
\usepackage{physics}
\usepackage{amsfonts}
\usepackage{tabularx}
\usepackage{bigstrut}
\usepackage{booktabs}
\usepackage{bm}
\usepackage{physics}

\newcommand{\dimension}{\ensuremath{\mathcal D}}

\journal{Annals of Physics}

\begin{document}

\begin{frontmatter}

\title{Topology and Many-body Localization}

%\author{Akshay Krishna, R. N. Bhatt}
%\address{Department of Electrical Engineering, Princeton University, Princeton NJ 08544, USA}


\author[mymainaddress]{R. N. Bhatt}
\ead{ravin@princeton.edu}

\author[mymainaddress,mysecondaryaddress]{Akshay Krishna}

\address[mymainaddress]{Department of Electrical Engineering, Princeton University, Princeton NJ 08544, USA}
\address[mysecondaryaddress]{KLA Corporation, 2350 Green Road Suite 100, Ann Arbor, MI, 48105, USA}

\begin{abstract}
We discuss the problem of localization in two dimensional electron systems in the quantum Hall (single Landau level) regime. After briefly summarizing the well-studied problem of Anderson localization in the non-interacting case, we concentrate on the problem of disorder induced many-body localization (MBL) in the presence of electron-electron interactions using numerical exact diagonalization and eigenvalue spacing statistics as a function of system size. We provide evidence showing that MBL is not attainable in a single Landau level with short range (white noise) disorder in the thermodynamic limit. We then study the interplay of topology and localization, by contrasting the behavior of topological and nontopological subbands arising from a single Landau level in two models – (i) a pair of extremely flat Hofstadter bands with an optimally chosen periodic potential, and (ii) a Landau level with a split-off nontopological impurity band. Both models provide convincing evidence for the strong effect of topology on the feasibility of many-body localization as well as slow dynamics starting from a nonequilibrium state with charge imbalance. 
\end{abstract}

\end{frontmatter}

%\linenumbers

\section{Introduction}

It took nearly two decades after Anderson’s seminal 1958 paper \citep{Anderson1958} establishing the existence of electron localization due to disorder to uncover the nature of the transition \citep{Wegner1976}, and thereby establish that all states in the non-interacting Anderson model with potential disorder were localized for dimensions two and below, using the scaling theory of localization \citep{GangofFour1979}. Since then, the study of Anderson localization has spawned a veritable industry, consisting of experiments, analytic theory and numerical approaches. What emerged was a beautiful story of the interplay of dimension and universality classes of Hamiltonians in the phenomenon of Localization. A small selection of review articles describing the progress in the twentieth century is in References \citep{Dyson1962, Lee1985, Altshuler1985, Milligan1985, MacKinnon1993, Belitz1994, Ohtsuki1999}. 

Soon after the scaling theory of localization \citep{GangofFour1979} was formulated, experiments on two-dimensional electron gases in semiconductors in a high magnetic field \citep{Klitzing1980} uncovered the phenomenon known as the integer quantum Hall effect (IQHE) – the quantization of the Hall conductance in two-dimensional systems in integer multiples of the fundamental unit $e^2/h$. The theoretical explanation which soon followed \citep{Laughlin1981}, showed that localization played a central role in the quantization of the Hall conductance over much of the phase diagram. However, the existence of a nonzero Hall conductivity pointed to a new result – that in such systems, the localization length diverged at certain critical energies \citep{Halperin1982}, in contrast to the situation in zero magnetic field. We now recognize this as being due to the topological nature \citep{Thouless1982, Arovas1988} of Landau levels, which have given rise to a much richer set \citep{Altland1997} of universality classes of disordered systems, beyond the original Wigner-Dyson classes \citep{Dyson1962}. This divergence has been much studied by diverse numerical methods \citep{Chalker1988, Huckestein1990, Huo1992, Slevin2009}, though the precise critical exponent \citep{Obuse2012, Zhu2019, Puschmann2019}, and even the nature \citep{Gruzberg2017, Bondesan2017} of the divergence is still being debated. 

The integer quantum Hall effect was followed a few years later by the even more surprising discovery \citep{Tsui1982} and theoretical understanding \citep{Laughlin1983} of the fractional quantum Hall effect (FQHE), where the quantization of the Hall conductance was at rational fractions (in units of $e^2/h$), and the explanation relied entirely on electron-electron interactions. The initial result of Laughlin \citep{Laughlin1983} was soon generalized to a hierarchy of fillings \citep{Haldane1983}; several field-theoretic approaches were attempted using the concept of composite particles, the most successful of which was the composite fermion approach of Jain \citep{Jain1989, Jain2007}. Jain’s approach gave rise to a hierarchy of FQHE states which was in amazing agreement\footnote{It should be noted that a small fraction of observed FQH states are not captured by the Jain sequence.} with experiment, both in identification, as well as strength (magnitudes of the excitation gap whose existence is responsible for the FQHE). The FQHE naturally gave rise to the question of localization due to disorder in such systems, such as would disorder destroy the topological nature of the ground state by closing the excitation gap? Though only a few quantitative numerical results exist to date, tracking both the topological character \citep{Sheng2003, Wan2005} and the entanglement entropy \citep{Liu2016, Liu2017} in the ground state shows that such a FQH Hall to non-topological insulator transition does take place. 

The phenomenon of spin localization was originally in Anderson’s mind when he wrote the 1958 paper \citep{Anderson1958}, motivated by the beautiful magnetic resonance experiments on doped semiconductors by Feher \citep{Feher1955, Feher1959b}. Research in spin localization saw significant progress in the last two decades of the twentieth century, motivated by experiments on magnetic properties – again in doped semiconductors \citep{Andres1981, Paalanen1986} as well as disordered quasi-1D organic salts \citep{Bulaevskii1972, Tippie1981}. Numerical and theoretical investigations \citep{Dasgupta1980, Bhatt1982, Bhatt1986, Milovanovic1989, Bhatt1992, Dobrosavljevic1992} led to the conclusion that disorder leads to spin localization at low energies, not only in one-dimension, but also in disordered electronic systems in higher dimensions. This could happen even when charge localization is absent, so the system is a metal at low temperatures with a diverging magnetic susceptibility over a finite range of the phase diagram!

The proper generalization of the story of localization to interacting electron systems came in the first decade of the twenty-first century, with the formulation and first true understanding of many-body localization. While several attempts had been made to generalize Anderson’s idea to interacting many-body systems, the crucial breakthrough \citep{Basko2006, Gornyi2005} came in the period 2005-6, particularly with the comprehensive paper by Basko, Aleiner and Altshuler \citep{Basko2006} using perturbative approaches. In a few years, complete (or infinite temperature) many-body localization (MBL) came to be understood \citep{Oganesyan2007, Pal2010} as a breaking of the eigenstate thermalization hypothesis \citep{Deutsch1991, Srednicki1994}, and the subject underwent an explosive expansion, see e.g.\ \citep{Serbyn2013, Huse2014, Kjall2014, Potter2015, Khemani2017, Dumitrescu2017}.
More details and references can be found in many excellent reviews \citep{Nandkishore2015, Altman2015, Abanin2017, Alet2018, Parameswaran2018}. Currently, MBL seems to be well established in random field spin chains, a one-dimensional spin model, whereas for spin models with randomness in higher dimensions ($d = 2$ and above), strong arguments have been made to support the contention that MBL is destroyed by rare fluctuation effects. Recent numerical studies of one-dimensional quasiperiodic systems \cite{Khemani2017b} show that MBL is more robust than in the disordered case, indirectly supporting the above contention.  Nevertheless, it should be recognized that because the Hilbert space grows exponentially with the total size ($\sim L^d$, where $L$ is the linear dimension), numerical studies, especially in $d > 1$, are limited to rather small sizes.  

Here we report the results of numerical studies of MBL in systems in the quantum Hall regime. Such studies are interesting on several fronts. First, the system possesses only the charge degree of freedom – the spin degree of freedom is quenched because of the high magnetic field; consequently, the Hilbert space does not grow as fast as in the case of fermions with both charge and spin degrees of freedom. Secondly, the model, while not one-dimensional, has the next lowest dimensionality ($d = 2$), thereby keeping the rate of growth of Hilbert space with size manageable. Thirdly, one can formulate it in a continuum version, and thereby avoid discrete lattice effects present in lattice models in $d > 1$. Finally, it allows one to study directly the effect of topology on MBL, something not studied in spin models. A direct motivation for studying this effect numerically comes from an earlier analytical study \citep{Nandkishore2014A} of MBL in Landau levels, which concluded that MBL was not possible in a Landau level broadened by disorder because of the diverging localization length at the single particle level. A first numerical study \citep{Geraedts2017B} confirmed that qualitative result, but found a much larger effect of topology than predicted by the analytic considerations.

\section{The quantum Hall regime and the lowest Landau level}

\subsection{The quantum Hall regime and the QH Hamiltonian}

We consider a generic system of $N$ electrons of mass $m$ and charge $-e$ moving in the $xy$ plane in the presence of a magnetic field, whose Hamiltonian is the sum of three terms: \begin{align}
    \hat{H} &= \hat{T} + \hat{U} + \hat{U}_\text{int} = \sum\limits_{i=1}^N \frac{\bm{\pi_i}^2}{2m} + \sum\limits_{i=1}^N V(\bm{r_i}) + \sum\limits_{i \neq j}^N V_\text{int}(\bm{r_i}, \bm{r_j}). \label{eq:Ham1}
\end{align}
Here $\bm{\pi} \equiv \mathbf{p} + e \mathbf{A}$ is the dynamical momentum operator, $V(\bm{r})$ is an arbitrary single-particle potential and $V_\text{int}(\bm{r}, \bm{r'})$ is the electron-electron interaction.
When the magnetic field $\bm{B}$ is constant and perpendicular to the plane, the free-electron part of the Hamiltonian turns into a sum over discrete Landau levels, each separated by the cyclotron energy $\omega_B \equiv eB / m$.
In the high-field limit, the Zeeman splitting and the cyclotron energy are both much larger than the potential terms, and the Hamiltonian above can be projected down to the lowest Landau level (LLL) as \begin{align}
    H &= \mathcal{P}_\text{LLL} \left[\hat{U} + \hat{U}_\text{int} \right] \mathcal{P}_\text{LLL}. \label{eq:Ham2}
\end{align}
where $\mathcal{P}_\text{LLL}$ is a projection operator to the LLL.
The kinetic energy of the electrons is completely quenched, and therefore the behavior of the system is completely governed by the interplay of disorder, applied potentials and interactions.
This is the situation we work with in the rest of the paper.
For computational reasons, we assume that the system has periodic boundary conditions, so the system lies on a torus with sides $L_x$ and $L_y$. 
The torus encloses an integer $N_\Phi$ of magnetic flux quanta, due to the magnetic field everywhere perpendicular to its surface.
For much of this paper, we further consider only square tori, so $L_x = L_y$. 
A natural length scale in the system is the magnetic length $l_B \equiv \sqrt{\hbar/eB}$, which we set to unity. One quantum of flux encloses an area $2 \pi l_B^2$.
In the following sections, the single-particle potential $\hat{U}$ includes both random disorder as well as a systematic term.

\subsection{Many-body localization and the $r$-statistic}

Exact diagonalization is a popular technique used in numerical studies of disorder and localization.
One of its great advantages is the ability to directly access the full spectrum of energies and spatial description of the wave functions, albeit for small sizes.
A telltale numerical signature of the onset of disorder is in the spacing between successive eigenvalues in the middle of the spectrum.
In an ergodic, or thermal phase, eigenvalues tend to repel each other, and the probability distribution $p(s)$ of the scaled spectral gaps is such that $p(s = 0) = 0$ and $P(s \to 0) \sim s^\beta$, where $\beta = 1, 2$ or $4$ depending on the universality class of the system.
On the other hand, in a localized phase, both in the single particle as well as many-body settings, the eigenvalues do not repel each other and spectral gaps follow a Poisson distribution, $p(s) \sim e^{-s}$.
Determining the full form of $p(s)$ from numerical simulation is not feasible currently in our problem due to the amount of ensemble averaging required, and more importantly due to finite size effects that lead to energy spacings of the order of $L^{-d}$.

A convenient single parameter used as a proxy to $p(s)$ to characterize these distributions is the mean level spacing ratio \cite{Oganesyan2007}, denoted by $r$. This quantity has been found to work well in the case of data that is limited to small sizes. Denoting the eigenvalue spacing between the $n$\textsuperscript{th} and $(n+1)$\textsuperscript{th} eigenstates by $\Delta E_n \equiv E_n - E_{n-1}$, $r$ is given by
	 \begin{align} r  = \langle r_n \rangle \equiv \left \langle \frac{\min(\Delta E_n, \Delta E_{n-1})}{ \max (\Delta E_n, \Delta E_{n-1} )} \right \rangle, \end{align}
where the angular brackets denote a combined average over both energies and realizations of disorder.
The energy resolved ensemble-averaged mean $r$ value gives clear signatures of the localization information of the underlying phase and has been used effectively in recent numerical studies of disordered many-body systems \cite{Pal2010, Cuevas2012, Johri2015, Luitz2015, Theveniaut2020, Giraud2020}.

In the context of the quantum Hall problem, at zero disorder, the eigenvalue statistics are governed by the Gaussian unitary ensemble (GUE), characteristic of an ergodic system with broken time-reversal symmetry.
In this case, $ r  \approx 0.5996$ \cite{Atas2013}.
In a disordered (localized) phase, the eigenvalue spacing distribution is Poissonian and $ r = 2 \ln 2 - 1 \approx 0.3862$.

\subsection{Absence of many-body localization in the lowest Landau level}

Our group first investigated the possibility of MBL in the LLL in 2017 \cite{Geraedts2017B} by calculating the dependence of the $r$ statistic on the strength of disorder for the Hamiltonian in Eq.\ \eqref{eq:Ham2}.
The single particle term was set to a real-space zero-mean Gaussian white noise disorder potential with strength $\langle V(\bm{r}) V( \bm{r'}) \rangle = W^2 \delta(\bm{r} - \bm{r'})$. Further, the interaction $\hat{U}_\text{int}$ considered was a pure $V_1$ Haldane pseudopotential \cite{Haldane1983}, which is more conveniently expanded in momentum space as \begin{align}
    V_\text{int}(\bm{k}) &= V_c (1 - k^2 l_B^2). \label{eq:pseudopot}
\end{align}
The strength of the interaction $V_c$ was fixed to unity, and set the energy-scale for the problem.
200 realizations each of systems of $N_e = 3, 4, 5, 6$ electrons  were diagonalized at filling $\nu = 1/3$, so that the number of flux quanta $N_\Phi$ was $9, 12, 15$ and 18 respectively (the side of the torus ranged from $\approx 7.5\ l_B$ to $\approx 10.6\ l_B$). 

\begin{figure}[t!]
    \centering
    \includegraphics[ width=0.48\columnwidth]{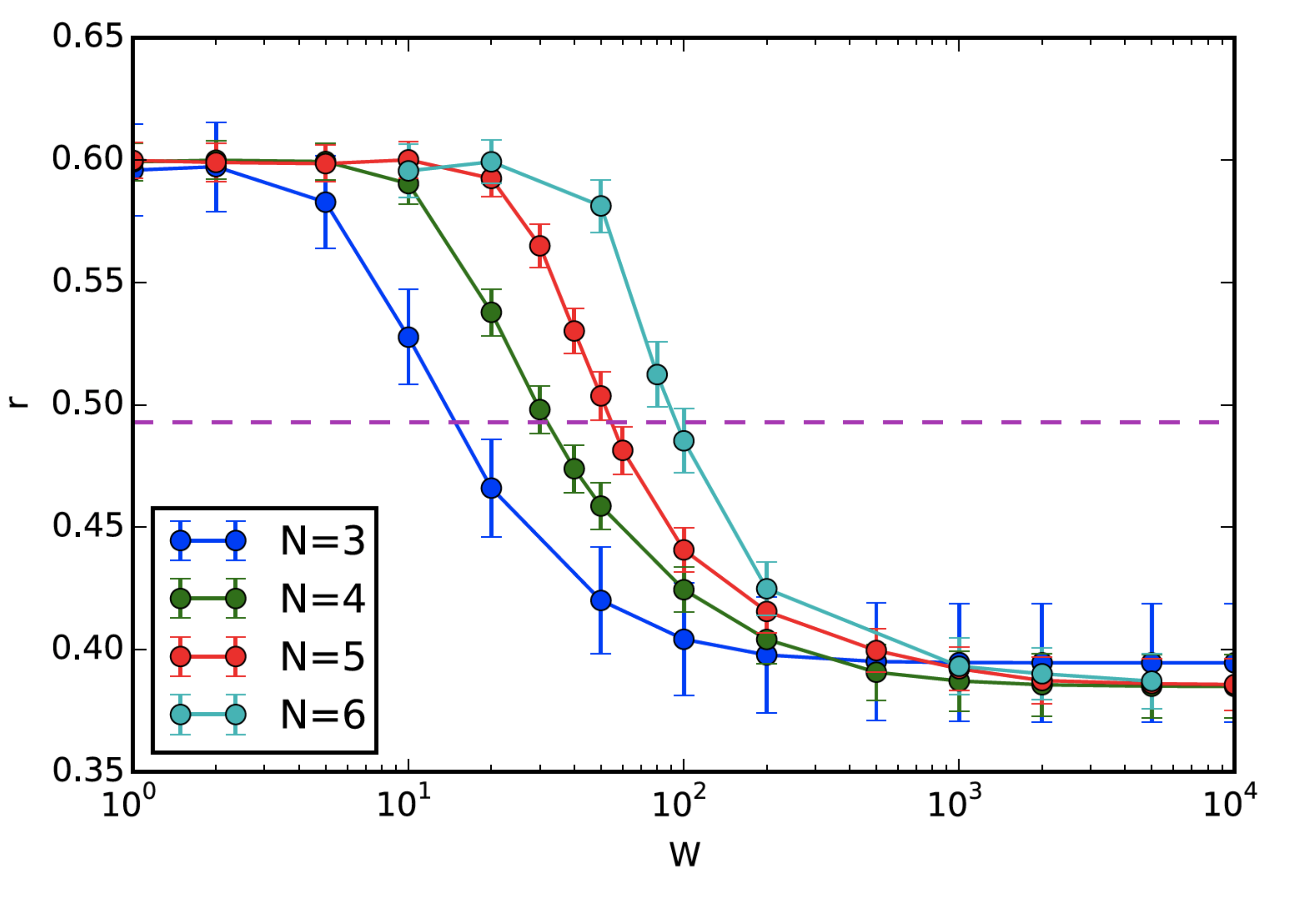}
    \includegraphics[ width=0.41\columnwidth]{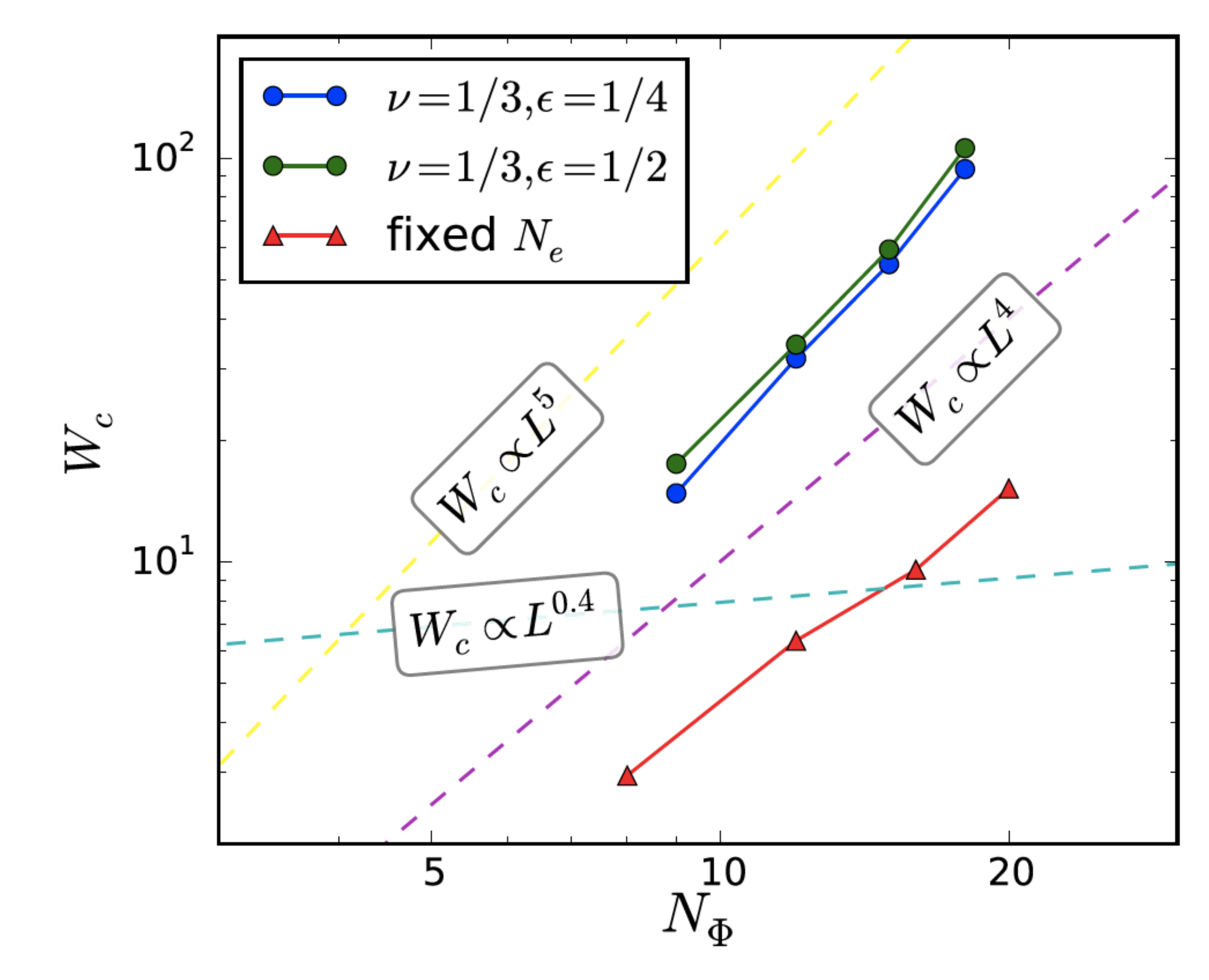}
    \caption{(Left) The variation of $r$ statistic with disorder strength $W$ in the lowest Landau level for four different system sizes at fixed filling $\nu = 1/3$. At low disorder, $r$ suggests thermal behavior, and at large disorder, it suggests localized behavior. However the critical disorder $W_c$ (i.e.\ the value of disorder at which $r$ attains a value midway between the GUE and Poisson values, indicated by the dashed horizontal line) increases rapidly with size, as is indicated by the x axis being linear in the logarithm of $W$.
    (Right) The critical disorder $W_c$ increases rapidly as a function of system size $N_\Phi$. Plots are made both at fixed filling and varying number of electrons (at two different energy windows $\epsilon$ in the spectrum), and at a fixed number of electrons and varying filling. In all cases, the critical energy seems to scale as a power law with large exponent $W_c \sim L^{4 -5}$. [Reprinted figure with permission from S.\ D.\  Geraedts and R.\ N.\ Bhatt, Absence of many-body localization in a single Landau level, Phys.\ Rev.\ B \textbf{95}, 054303 (2017).  Copyright (2017) by the American Physical Society.]}
    \label{fig:Ger1}
\end{figure}

The findings shown in Fig.\ \ref{fig:Ger1} seem to emphatically reject the possibility of MBL in this system.
As is clear from the figure, the critical disorder $W_c$ (measured as the value needed to reach an $r$
value of $0.493$, midway between the result for GUE and Poisson statistics) rapidly increases with
the size $L$, consistent with $W_c \to \infty$ as $L \to \infty$.
In fact, a few years earlier Nandkishore and Potter
\cite{Nandkishore2014A} had argued that the divergence of the single particle localization length within the
disorder broadened Landau level in the non-interacting system would lead to a destruction of
MBL in the interacting system in the thermodynamic limit.
However, their arguments suggested that the critical disorder strength would scale as $W_c \sim L^{\sim 0.4}$, whereas the numerical results gave an exponent that was an order of magnitude larger. This unexpected result provides yet another example of the surprising fragility of MBL.

It should be remarked that unlike the wide differences in the ground state of the interacting system at different filling fractions $\nu$, the filling fraction does not seem to matter for the MBL problem.
This is consistent with the fact that MBL is an infinite temperature transition corresponding to all states, not just the ground state. It is also evident from the much larger
values of the disorder parameter $W_c$ than the ground to first excited state gap in the clean system for any $\nu$.

\section{Models of Landau level subbands}

These results beg the question -- what is the primary cause for the absence of MBL in the LLL? Is it primarily topology, or dimensionality? Would a 2-D system with no topological character be more amenable to an MBL phase at large disorder?
In this context, we tried to uncouple the peculiar non-zero Chern character of the Landau level wave functions from their 2-D nature in a series of studies \cite{Krishna2019, Krishna2019B}. 

The aim was to split the LLL into subbands, each subband having a separate Chern character.
This was accomplished by adding a single particle potential to the Hamiltonian in two different ways.

\subsection{Periodic potential \label{subsec:cleanlattice}}

We first consider adding an arbitrary single particle periodic potential $V(x,y) = V(x+a, y) = V(x, y+b)$ on a rectangular unit cell of size $a \times b$ to the disorder potential $V_\text{dis}(\bm{r})$, so that :
\begin{align}
V(\mathbf{r}) = V(x, y) + V_\text{dis}(\bm{r}).
\label{eq:cleanpotential}
\end{align}
This is known as a Hofstadter potential \cite{Hofstadter1976} with disorder. 
In terms of a Fourier series expansion, \begin{align}
V(x, y) = \sum\limits_{m_x, m_y} v_{m_x, m_y} e^{i 2 \pi (m_x x/a + m_y y/b)} \;. \label{eq:Fourier}
\end{align}

At zero disorder, the single particle term possesses discrete translation symmetry, defined by the unit cell of the potential.
However, due to the peculiar non-commuting nature of magnetic translation operators \cite{Zak1964}, the \emph{magnetic unit cell}, distinct from the unit cell of the potential, must enclose an integer number of flux quanta. 
If the potential unit cell encloses $\frac{p}{q}$ flux quanta (where $p$ and $q$ are co-prime integers), a magnetic unit cell with sides $qa \times b$ may be defined, which is a repeating building block of the entire system.
Then, the usual procedure of Bloch's theorem is applied in order to obtain the band structure of the system.
The number of these \emph{Hofstadter subbands} is $p$ \cite{Hofstadter1976}.
At a fixed $\bm{k}$, the band structure can be calculated by diagonalizing a $p \times p$ matrix, whose elements depend on the Fourier coefficients $v_{m_x, m_y}$ as well as the Gaussian form factor and complex phase factor of the LLL \cite{Krishna2019B}.
The Chern numbers $C$ of each of the $p$ subbands must obey the diophantine equation \cite{Thouless1982} $pC + qs = 1$, $C, s \in \mathbb{Z}$.
We set $p=2, q=1$ for concreteness, giving us one topological sub-band with $C=1$, and another non-topological sub-band with zero Chern number.

Our goal is to obtain widely separated bands with small dispersions $E_b$ and a large gap $E_g$, so that disorder $V_{\text{dis}}$ and interaction $V_{\text{int}}$ can lie in an intermediate range, $E_b \ll V_{\text{dis}}, V_{\text{int}} \ll E_g$.
We therefore optimize the periodic potential for maximal flatness $E_b / E_g \to 0$ in the space of square-symmetric potentials, with $|m_x|, |m_y| < 4$.
%This results in the Fourier coefficients shown in Table \ref{tab:coefficients} and a real space potential as plotted in Fig.\ \ref{fig:potential}(a).
The overall normalization (irrelevant to the $E_b/E_g$ ratio) is chosen to yield unit bandwidth $E_b = 1$.
%Given the absence of terms with both $m_x$ and $m_y$ even, the resulting band structure has the symmetry $E_1(\mathbf{k}) = -E_2(\mathbf{k})$.
Our optimal choice of Fourier coefficients \cite{Krishna2019B} yields a remarkably large bandgap-to-bandwidth ratio of $E_b/E_g \approx 8735$, allowing us to tune disorder and interaction over several orders of magnitude while safely neglecting inter-subband mixing. Another interesting feature of our choice of Fourier coefficients is that the two bands have an exact symmetry $E_1(\bm{k}) = -E_2(\bm{k})$. The two subbands are energetically indistinguishable, but have different Chern character.

\begin{figure}[t!]
    \centering
    \includegraphics[ width=0.44\columnwidth]{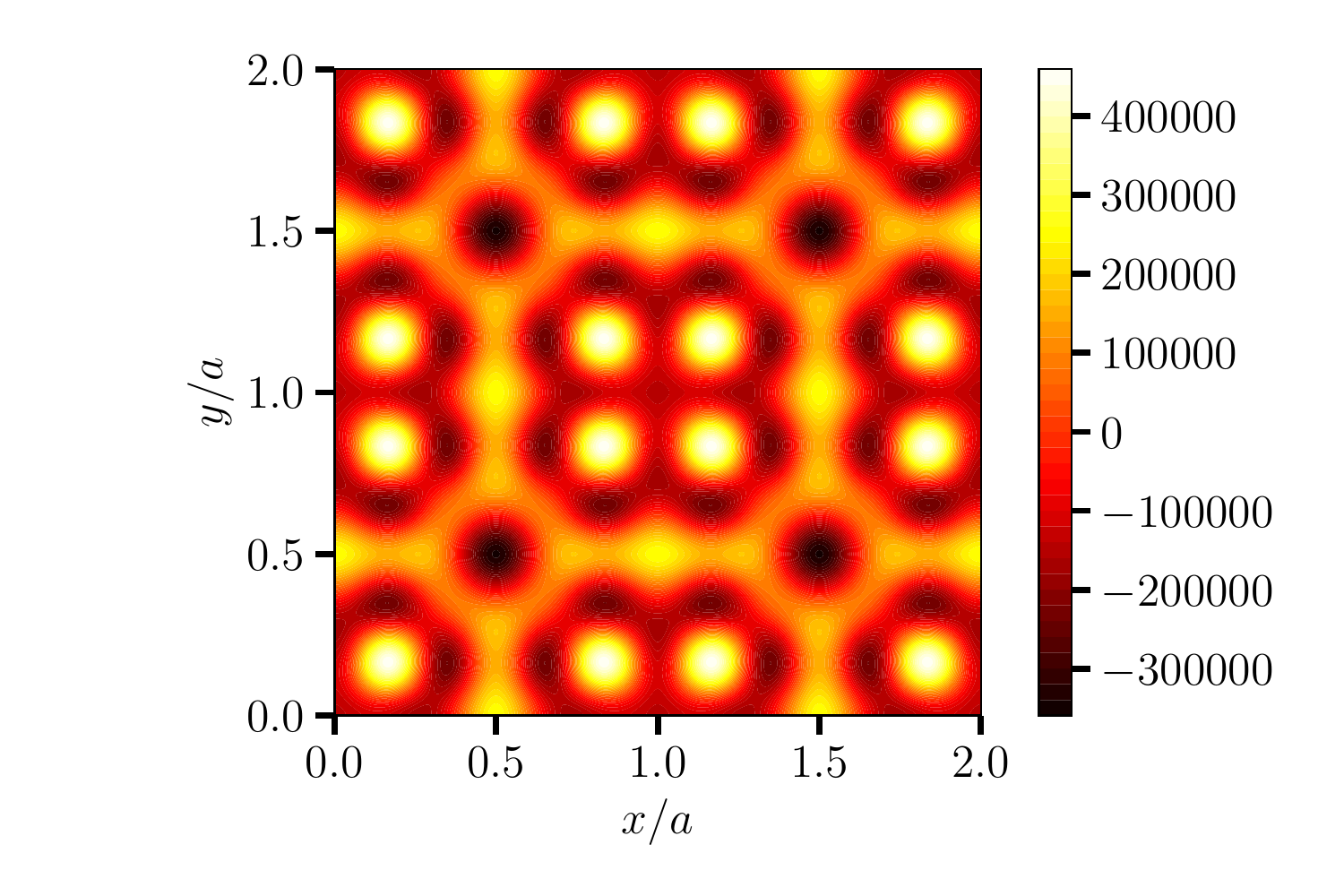}
    \includegraphics[ width=0.49\columnwidth]{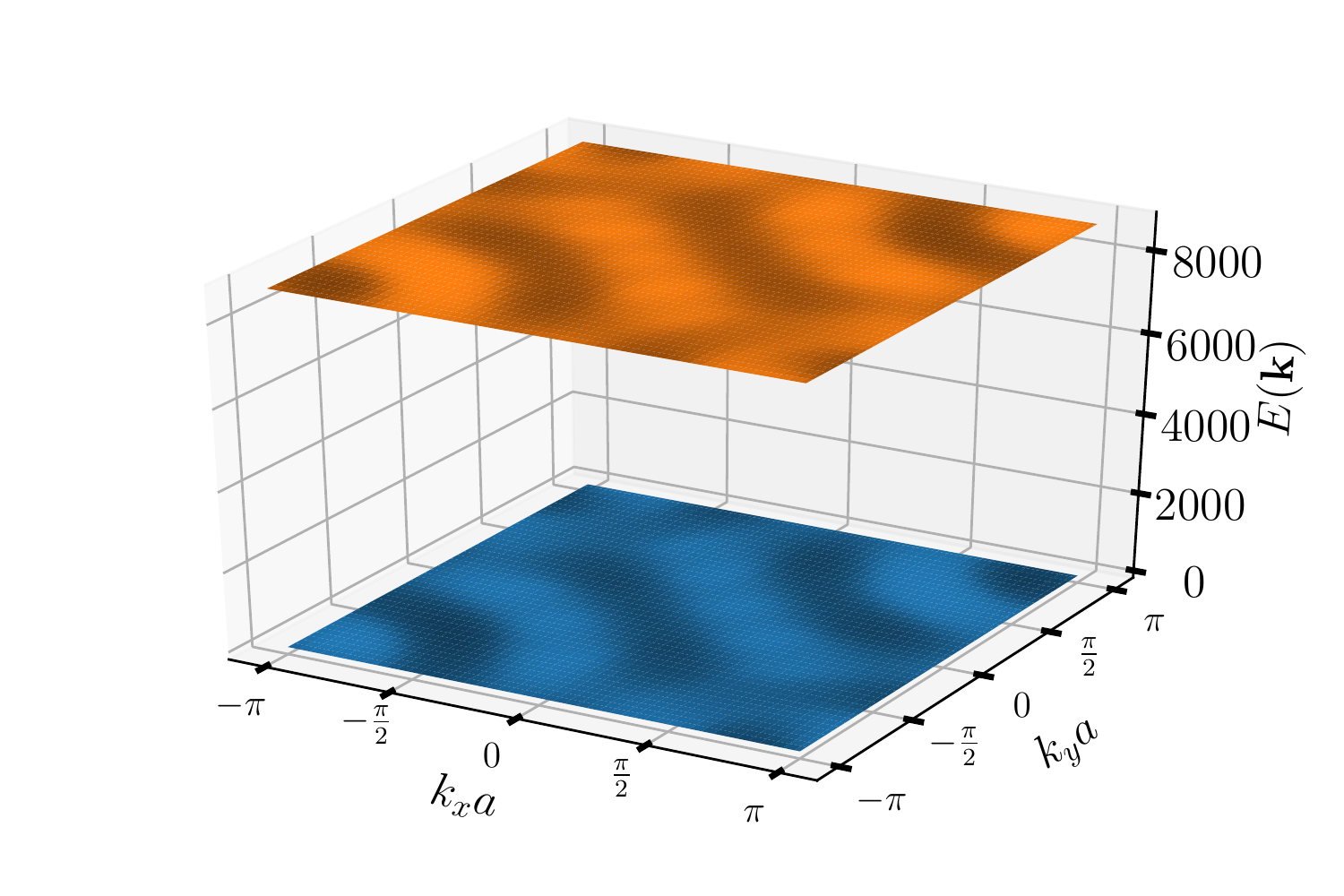}
    \caption{(Left) The periodic potential of Eq.\ \eqref{eq:Fourier} is plotted on a plaquette of $2 \times 2$ unit cells in real space.
    The coefficients $v_{m_x, m_y}$ are optimized for high flatness. We set the unit cell to be square, so that $a = b$, and constrain the coefficients so that they maintain square symmetry. See Ref.\ \cite{Krishna2019B} for details of the coefficients.
    (Right) The dispersion $E(\bm{k})$ in the Brillouin zone is plotted for both Hofstadter subbands. The two subbands are exact mirror opposites of each other, and have a bandwidth $E_b$ of 1 each. The band gap $E_g$ is $\approx 8735$.}
    %[Reprinted figure with permission from A.\ Krishna, M.\ Ippoliti and R.\ N.\ Bhatt, Many-body localization in Landau-level subbands Phys.\ Rev.\ B \textbf{99}, 041111(R) (2019).  Copyright (2019) by the American Physical Society.]}
    \label{fig:flat1}
\end{figure}
%%%%%%%%%%%%%%%%%%%%%%%%%%%%%%%%%%
%%%%%%%%%%%%%%%%%%%%%%%%%%%%%%%%%%

\subsection{Impurity band \label{subsec:impurelattice}}

A special case of the periodic potential consists of a lattice of point-like impurities, modeled by delta functions \cite{Ippoliti2018}.
\begin{align}
V(\mathbf{r}) = -V_0 \sum\limits_{\bm{t}} \delta(\mathbf{r} - \bm{t}), \label{eq:deltapotential}
\end{align}
where $\bm{t}$ is a lattice vector. This lattice could be non-rectangular in general.

As shown by Prange \cite{Prange1981}, a single delta function splits a single localized bound state from the
Landau level, leaving the energies of the rest of the states unchanged at $E = 0$, the LLL energy.
This can be understood as the wavefunctions have $N_\Phi$ nodes (zeroes), and the eigenstates with
unchanged energies have a zero at the delta function. For multiple delta functions, one gets an
$E = 0$ level with degeneracy reduced by the number of delta functions, assumed to be smaller than the number of flux quanta,
plus an ``impurity band'' away from E = 0, made up of eigenstates that do not have a zero at
every delta function potential.

Coming back to our delta-function lattice, we consider the case when the total number of delta functions $N_\delta$ is such that $\frac{N_\phi}{N_\delta} = \frac{p}{q}$, for coprime integers $p$ and $q$.
If $N_\delta < N_\phi$, we obtain $p - q$ degenerate bands at $E = 0$.
This manifold is topological with total Chern number $C=1$.
The remaining $q$ `\emph{split-off}' bands are non-topological and are centered around energy $-V_0$.

In order to incorporate the effects of disorder, we randomize the strengths and positions of scatterers in the lattice of point impurities by replacing Eq.\ \eqref{eq:deltapotential} with \begin{align}
%V_{\text{1-body}}(\mathbf{r}) = - \sum\limits_{n_1, n_2} V_{n_1, n_2} \delta(\mathbf{r} - n_1 \mb{a}_1 - n_2 \mb{a}_2). \label{eq:deltapotential_dis}
V(\mathbf{r}) = - \sum\limits_{n=1}^{N_\delta} V_{n} \delta(\mathbf{r} - \mathbf{r}_n). \label{eq:deltapotential_dis}
\end{align}

The advantage of the delta function potential is that we can introduce both positional disorder
by placing them at random positions instead of a lattice $\{ \bm{t} \}$, as well as magnitude disorder by
choosing random coefficients instead of a constant $V_0$. The former implies that there are no additional
lattice commensuration conditions (for $d > 1$) beyond the requirement that the sample contain
an integer number of electrons (as in one dimension).

We choose $V_{n}$ to be independently and identically distributed uniform random variables in $[1-W, 1 + W]$, where we restrict $0 < W < 1$.
This choice fixes the gap between the split-off states and the rest of the degenerate LLL to be of the order of 1.
The positions of the scatterers $\mathbf{r}_n$ are randomly distributed on the torus, with a circular exclusion zone around each scatterer of area $2 \pi l_B^2 \frac{N_\phi}{N_\delta} \rho$. 
Similar to the lattice case, if $N_\delta < N_\phi$, there is a manifold of $N_\phi - N_\delta$ degenerate states at zero energy with total Chern number $C=1$. 

\begin{figure}[t!]
\centering
\includegraphics[trim={0 0.0cm 0 0.0cm}, clip, width=0.6\textwidth]{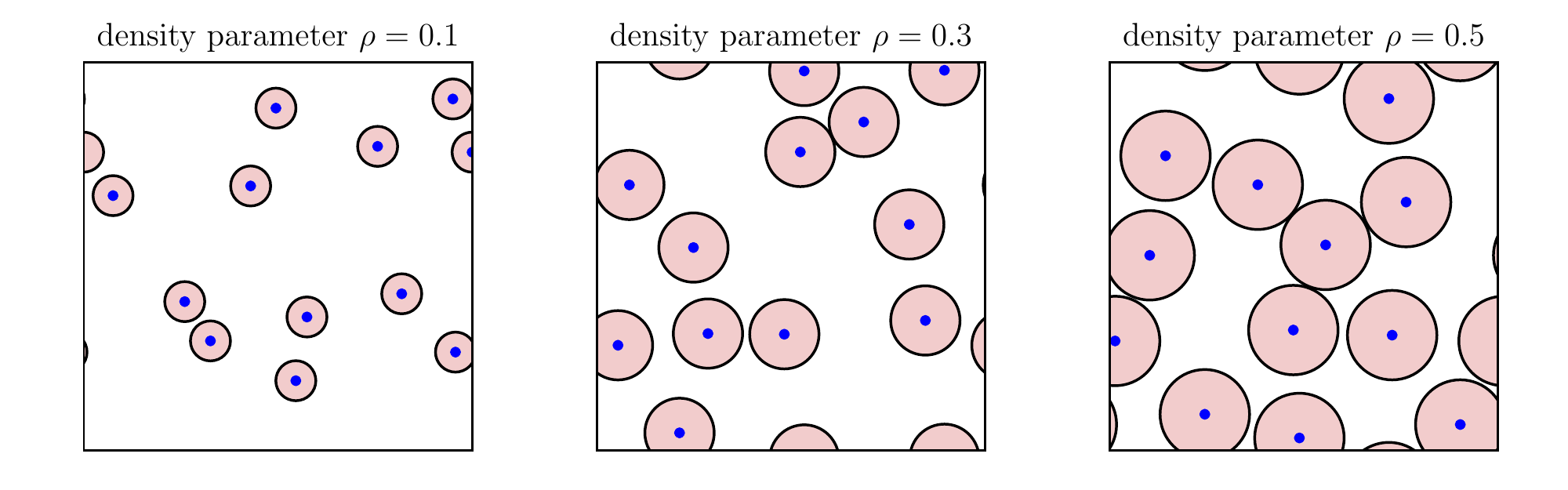}
\caption{For the impurity band model, point impurity scatterers (blue dots) are randomly distributed across the system. Here we show examples of the positions of $N_\delta = 12$ scatterers on a square torus for three different values of the density parameter $\rho$. The $\rho$ parameter is the fraction of the total area included in the circular red exclusion zones.  Increasing $\rho$ leads to reduced positional randomness, and avoids configurations where the scatterers can be densely concentrated in small patches while leaving others empty.
\label{fig:dens}}
\end{figure}

The width of the remaining $N_\delta$ split-off states is controlled by the disorder, which has two independent components.
The randomness in the scatterers' strengths is controlled by $W$, and randomness in their positions is controlled by the density parameter $0 < \rho < \frac{\pi}{2 \sqrt{3}} = 0.907$.
The upper bound for $\rho$ comes from a triangular lattice, which is the closest possible packing in two dimensions.
For large $\rho$, the distribution of scatterers becomes more regular (the maximum value indeed forces the configuration to be a triangular lattice with no randomness left). 
At the opposite end, $\rho=0$ corresponds to maximal randomness and allows two scatterers to sit arbitrarily close to each other, thus entirely closing the band gap. 
The effect of changing $\rho$ can be visualized in Fig.\ \ref{fig:dens}.

%(if two scatterers sit at the same exact position, then $N_\delta \mapsto N_\delta-1$, and the split-off band loses one state).

%1.	Adding the periodic lattice potential; splitting Landau level into 2 or more bands, Chern numbers. Optimizing the potential for flatness. Can include figures of band dispersion and 

%2.	The continuum model: creating an impurity band from the Landau level. LL retains C = 1, while impurity band is non-topological (C = 0). Avoiding rare fluctuations by restricting rare configurations (hard-core repulsion). See slide referring to Prange, 1981 and to Brezin, Gross and Itzykson, 1983. 

\subsection{Overview of the two models}

The two models each have relative advantages, as well as corresponding disadvantages.
In the periodic potential case of Sec.\ \ref{subsec:cleanlattice}, it is possible to make the subbands extremely flat by tuning the Fourier coefficients of the potential, and thus enable further projection of the disordered and interacting Hamiltonian from the LLL to a single subband of choice. The Fourier coefficients of the periodic potential can be contrived such that two subbands have identical and opposite dispersions $\pm E(\bm{k})$. The only way to distinguish them is through their topological character.
However, the constraint of putting the potential on a unit cell imposes severe restrictions due to periodic boundary conditions. Only a small number of system sizes  can be simulated.

The impurity band model of Sec.\ \ref{subsec:impurelattice} suffers from no such restrictions, as the delta functions are free to move around. It is therefore possible to simulate more sizes and do more effective finite-size scaling in this model.
However, an ensemble over the positions has to be performed, leading to longer computation times.
Also, the bandgap-to-bandwidth ratio $E_b/E_g$ of this system is much smaller than for the periodic potential. 
Finally, one cannot reach the zero-disorder (GUE) limit as even with $W=0$, there is positional disorder present. 
This constrains us to a smaller range of disorder and interaction strengths before inter-subband mixing renders the projection to the nontopological split-off band unphysical.

Nevertheless, the complementary advantages offered by the two models helps significantly in interpreting the data even with the small sizes that can be simulated.
%3.	Relative advantages and disadvantages of two models: (a) in first model can make bandwidth very small, can make both subbands very comparable in terms of bandwidth, but with differing topological character. (b) in continuum model not constrained to sizes characterized by integer linear dimensions, only integer total number – like in 1d spin chains, so possible to do more sizes.

\section{Results}

\subsection{Disorder in the non-interacting system}

We first briefly comment on the nature of single-particle localization in the periodic potential model of Sec. \ref{subsec:cleanlattice}.
In this model, we set the disorder to be a random short-range Gaussian correlated potential of the form \begin{align}
    \langle V_\text{dis}(\bm{r}) V_\text{dis}( \bm{r'}) \rangle = W^2 \sigma^{-2} e^{-\left| \bm{r} - \bm{r'} \right|^2 / 2 \sigma^2 } \label{eq:correldis}
\end{align}
The length scale $\sigma$ is a tunable length scale parameter.
We found that this type of disorder led to more localized wave functions than the delta-correlated disorder \cite{Krishna2019B}.

In order to quantify the spatial footprint of the wave functions, we use the inverse participation ratio $P_2 = \int d^2 \bm{r} |\psi(\bm{r})|^4$, where the wave function $\psi(\bm{r})$ is normalized.
A localization length is defined as $\xi \equiv 1 / \sqrt{2 \pi P_2}$. 
This choice ensures that a wave function of the form  $\psi(\bm{r}) = e^{-r/\xi}$ has localization length $\xi$.
The mean localization length is calculated by ensemble averaging over at least 200 disorder realizations for every combination of system parameters that we study.
Since it depends on energy, we perform the ensemble averaging within small energy windows, to resolve $\xi(E)$ across the entire spectrum.
By the scaling theory of localization, if there is any disorder at all in a 2-D system without magnetic field, $\xi(E)$ is expected to be finite for all energies. However, in the presence of a magnetic field, $\xi(E)$ diverges at the center of the Landau level, as is well known. More generally, a system with non-zero Chern number $C$ is expected to have $C$ critical energies at which the localization length diverges in the presence of disorder.

\begin{figure}[ht!]
    \centering
    \begin{minipage}{0.49\columnwidth}
    \includegraphics[ scale=0.370]{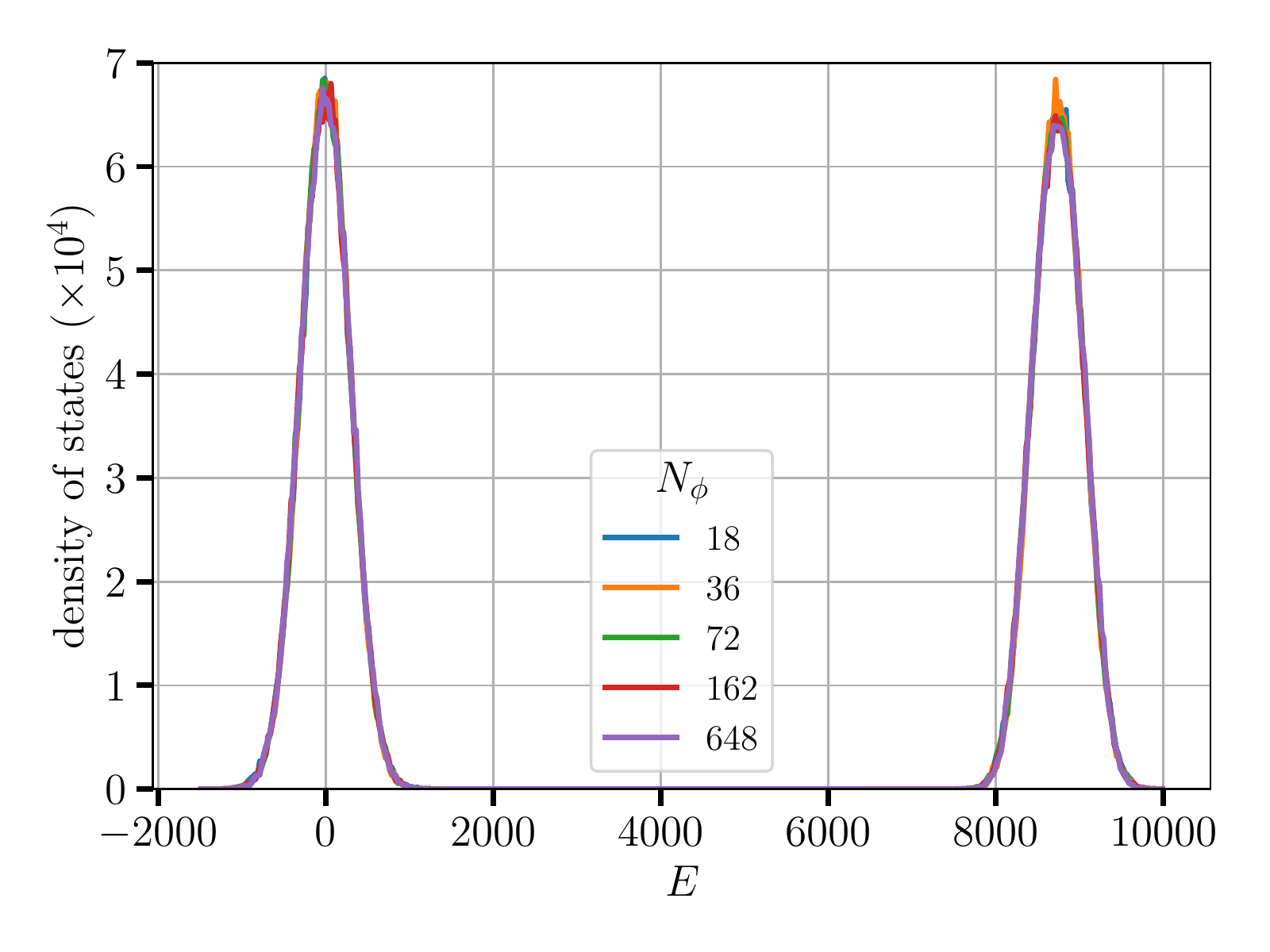}\\
    \includegraphics[ scale=0.370]{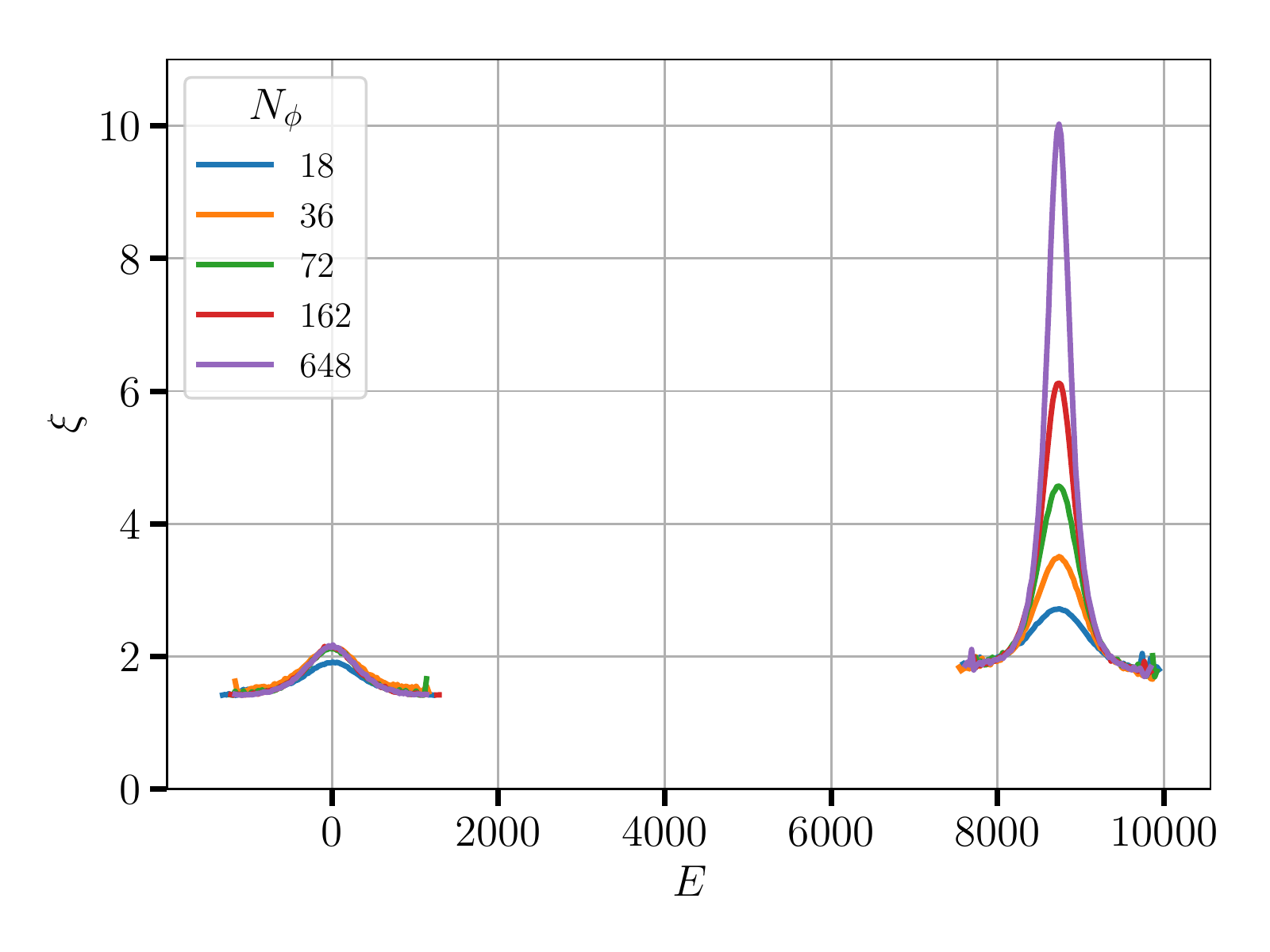}
    \end{minipage}
    \begin{minipage}{0.49\columnwidth}
    \includegraphics[ scale=0.5]{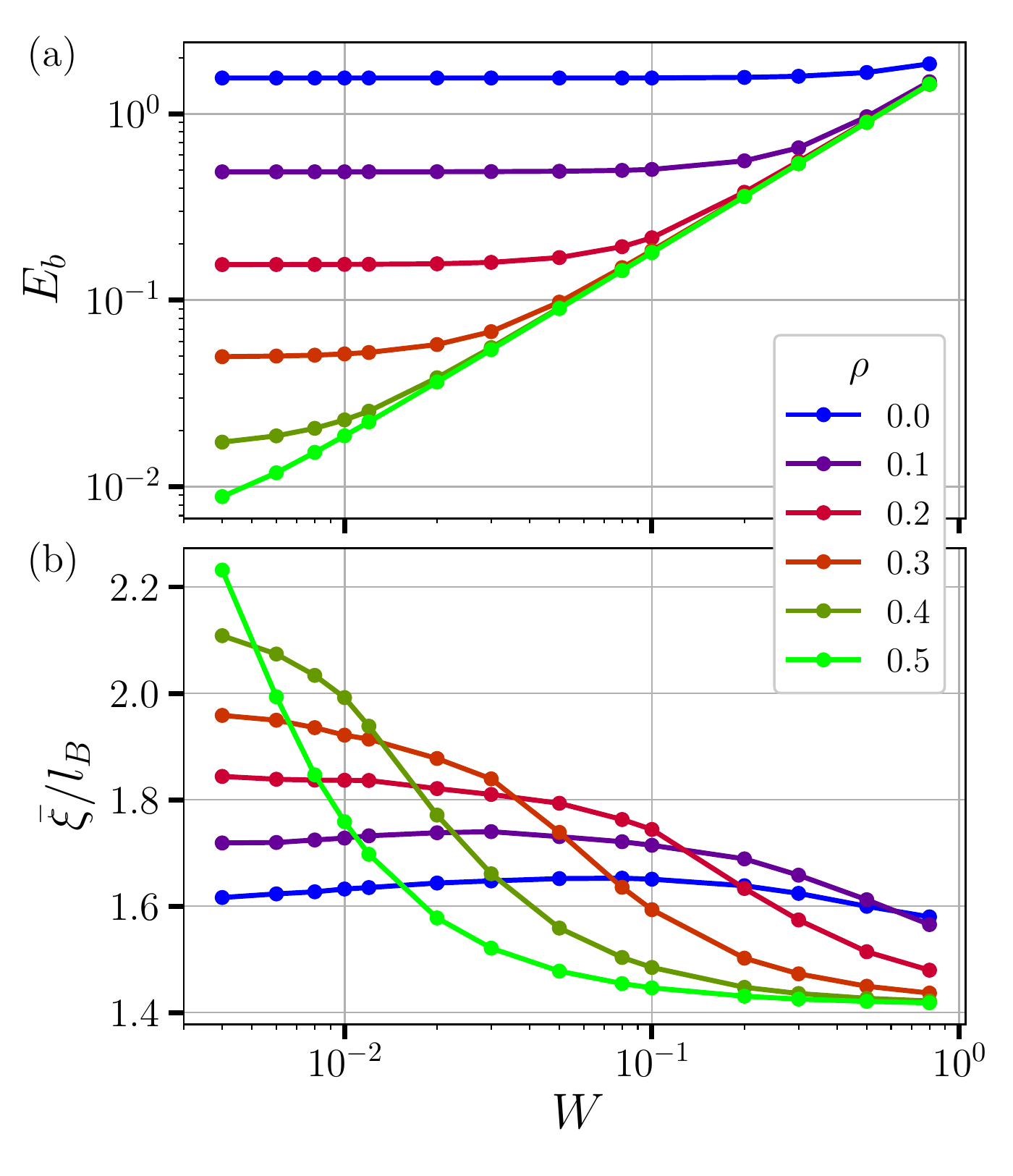}
    \end{minipage}
    \caption{(Left top) The single-particle density of states of the periodic potential model in Eq.\ \eqref{eq:cleanpotential}. We use the same flat-band potentials as shown in Fig.\ \ref{fig:flat1}, but additionally add correlated disorder, Eq.\ \eqref{eq:correldis}, with width $\sigma = l_B$ and strength $W = 500$. The disorder is much larger than the bandwidth ($=1$), and causes the bands to spread out, but is much smaller than the band gap so they don't mix. We simulate square tori of five different sizes $N_\Phi$ and find that the densities of states are well-converged, and the densities of states of the two subbands look very similar to the naked eye.
    (Left bottom) The mean localization length $\xi$ for the same system, in units of magnetic length $l_B$.  We measure energies relative to center of the $C=0$ band in the absence of disorder. The non-topological band with $C=0$ remains centered around zero energy and has a short localization length of $\approx 2$, which seems well-converged for all system sizes. The topologically nontrivial $C=1$ band on the centered at $E \approx 8735$ has a diverging localization length at its center, which can be clearly seen in these finite-size simulations.
    (Right) The bandwidth $E_b$ and average localization length $\bar{\xi}$ of the split-off states for the impurity-band model described in Eq.\ \eqref{eq:deltapotential_dis} is plotted as a function of amplitude disorder $W$ for six different values of the density parameter $\rho$. The bandwidth $E_b$ is defined as the energy interval within which 90\% of the split-off states lie. Recall that in this model, the band gap is normalized to 1. We set $N_\phi = 6 N_\delta$ and the system size to $N_\phi = 3000$.
    [Reprinted figure with permission from A.\ Krishna, M.\ Ippoliti, and R.\ N.\ Bhatt, Localization and interactions in topological and nontopological bands in two dimensions, Phys.\ Rev.\ B \textbf{100}, 054202 (2019).  Copyright (2019) by the American Physical Society.]}
    \label{fig:Bha11}
\end{figure}

In left half of Fig.\ \ref{fig:Bha11}, we show how this comes to pass in the context of the periodic potential model -- the localization lengths behave dramatically differently in the two subbands, even though the densities of states look very similar. 
The localization length in the nontopological subband is of the order of $2 l_B$, which is very small, and luckily for us, well below even the small system sizes that we are able to simulate in the case of the many-body problem. 
In contrast, the $C=1$ subband shows clear signs of a divergent localization length, just like the full LLL.
This dichotomy portends possibly different responses to MBL when interactions are turned on in the next section.

In the impurity band model of Sec.\ \ref{subsec:impurelattice} (see the right side of Fig.\ \ref{fig:Bha11}), we see that a reasonable value for the disorder parameters $W$ and $\rho$ leads to a small and well-converged localization length (of the order of $2 l_B$) for the split-off states well before the disorder is strong enough to close the gap.

%\akshay{Part A – Non-interacting part Comparing the localization lengths in two subbands of model 1.}

\subsection{Disorder in the interacting system and the search for MBL}
%\akshay{Part C – Charge imbalance equilibration in Model 1.}

Having demonstrated in two ways that we can reliably extract disordered well-separated subbands of desired topological character from the LLL, we now proceed to turn on interactions and attack the full many-body problem.
We continue to use the pure Haldane $V_1$ pseudopotential interaction as in Eq.\ \eqref{eq:pseudopot}.

The Hamiltonian we now work with is a modified version of Eq.\ \eqref{eq:Ham2}, because we additionally project it further, \begin{align}
    H &= \mathcal{P}_\text{sb} \mathcal{P}_\text{LLL} \left[\hat{U} + \hat{U}_\text{int} \right] \mathcal{P}_\text{LLL} \mathcal{P}_\text{sb}. \label{eq:Ham33}
\end{align}
$\mathcal{P}_\text{sb}$ is a projector to a desired disrodered subband of our choice -- either of the two subbands in the periodic potential model, or the split-off states in the impurity band model.

We investigate the onset of a possible many-body localization transition by performing exact diagonalization on systems of small sizes. The addition of interaction brings in an additional energy scale into the problem, so now there are two independent parameters: the interaction strength $V_c$, and the disorder strength characterized by $W$ (and $\rho$, in the impurity band model).
For the periodic potential model, the bandwidth and band gap of the single particle subbands are $1$ and $\approx 8735$ respectively, as mentioned previously.
We fix $V_c = 8$ and vary the disorder $W$. We found that at and above this value of $V_c$, the results seemed to depend largely only on the ratio $W/V_c$, thus simplifying the analysis.
In terms of the geometry, we choose square tori with integer flux quanta commensurate with the constraints of the lattice. 
In order to accommodate sizes that do not just have a perfect square number of unit cells, we allow rotate the torus with respect to the lattice. This allows us to simulate 6 different sizes between $N_\Phi = 16$ and $N_\Phi = 36$ flux quanta.
Our investigation covers a range of fillings $\nu$ between zero and half of the desired subband, with an emphasis on $\nu = 1/3$.
As shown earlier by Geraedts and Bhatt \cite{Geraedts2017B}, the filling fraction expectedly plays no significant role in this analysis -- since MBL is an infinite temperature transition, does not depend strongly on the properties of the ground state.

For the impurity band model, the energy scales are somewhat different. 
The gap between the split-off bands and the rest of the unperturbed LLL is of order 1. We set the number of scatterers $N_\delta = N_\Phi/6$, fix the density parameter such that the band-width $E_b \approx 0.1$, and the interaction $V_c = 1$. 
This density parameter $\rho = 0.3$ turns out to be far from the maximum while being large enough so rare configuration effects are minimized, unlike Poisson-distributed point particles.
Then we vary the amplitude of disorder $W$ and probe for the possibility of MBL.
In this case we are able to simulate somewhat larger sizes, ranging from $N_\Phi = 54$ through $N_\Phi = 108$, corresponding to a torus of side between $\approx 18.4\ l_B$ and $\approx 26\ l_B$.

Exact diagonalization yields both the eigenvalues, as well as the eigenvectors of the system.
The $r$ statistic uses the eigenvalues as a diagnostic for localization.
Motivated by experimental methods used in optical lattices and cold atoms, we have also examined long-time evolution from a non-equilibrium state as yet another metric for localization.
In many experiments on MBL \cite{Schreiber2015, Smith2016, Choi2016, Neyenhuis2017, Luschen2017, Luschen2017B, Kohlert2019}, the system is initially prepared in a non-equilibrium state,  with a spatial density inhomogeneity.
In the ergodic phase, unitary time evolution from any initial state (including an inhomogeneous one) should scramble the system completely, and wipe away the memory of the initial state.
On the other hand, when MBL kicks in, memory of initial conditions is retained to arbitrarily long times under unitary evolution, so a finite residual imbalance should be observed even after infinite time.

In order to probe this effect numerically, we consider the operator
\begin{align}
\hat{M} = \int \mathrm{d}r \ c^\dagger_{\bm{r}} c_{\bm{r}} \cos (2 \pi x / L_x)\;.
\end{align}
We initialize an almost homogeneous state with a small $\hat{M}$ perturbation, such that the initial density matrix is $\hat{\rho}_{0} = \frac{1 + \epsilon M}{\dimension}$, 
where $\epsilon$ is a small positive coefficient and $\dimension$ is the Hilbert space dimension.
This is very close to an infinite temperature state, with equal weight on all states in the spectrum.
The initial amplitude of charge density imbalance is\begin{align}
M_0 &= \Tr \left[ \hat{\rho}_{0} \hat{M} \right] = \frac{\epsilon}{\dimension} \Tr \hat{M}^2 =  \frac{\epsilon}{\dimension} \sum\limits_m \expval{\phi_m | \hat{M}^2 | \phi_m},
\end{align}
where the sum is over all eigen states.
This means the system initially will have a small cosine charge density imbalance along the $x$ direction.
Unitary evolution will relax this imbalance in the ergodic phase but not completely in the MBL phase.

The amplitude of the charge density imbalance at times $t>0$ depends on the time-evolved density matrix
\begin{equation}
\hat{\rho}_t = e^{-iHt} \hat{\rho}_0 e^{iHt} = \sum_{m,n} e^{i (E_n-E_m)t} | \phi_m \rangle \langle \phi_m | \rho_0 | \phi_n \rangle \langle \phi_n | \;.
\end{equation}
At long times $t \gg \hbar / \delta E$ (where $\delta E$ is the typical many-body energy spacing), the off-diagonal density matrix elements accumulate phases that time-average to zero (we assume the disorder prevents any degeneracies).
In this limit, denoting $\lim_{T\to\infty} \frac{1}{T} \int_0^T dt\ x(t)$ by $x_\infty$ , we have
\begin{equation}
\hat{\rho}_{\infty} = \sum_m ( \langle \phi_m | \hat{\rho}_0 | \phi_m \rangle )  | \phi_m \rangle  \langle \phi_m |
\end{equation}
and therefore
 \begin{align}
 M_\infty  &= \Tr \left[ \hat{\rho}_{\infty} M \right] 
= \frac{\epsilon}{\dimension} \sum\limits_m \left[ \expval{\phi_m | M| \phi_m} \right]^2.
\end{align} 

\begin{figure}[ht!]
\centering
\centering
    \begin{minipage}{0.61\columnwidth}
    \includegraphics[trim={0 0.0cm 9.1cm 0.0cm}, clip, width=0.99\textwidth]{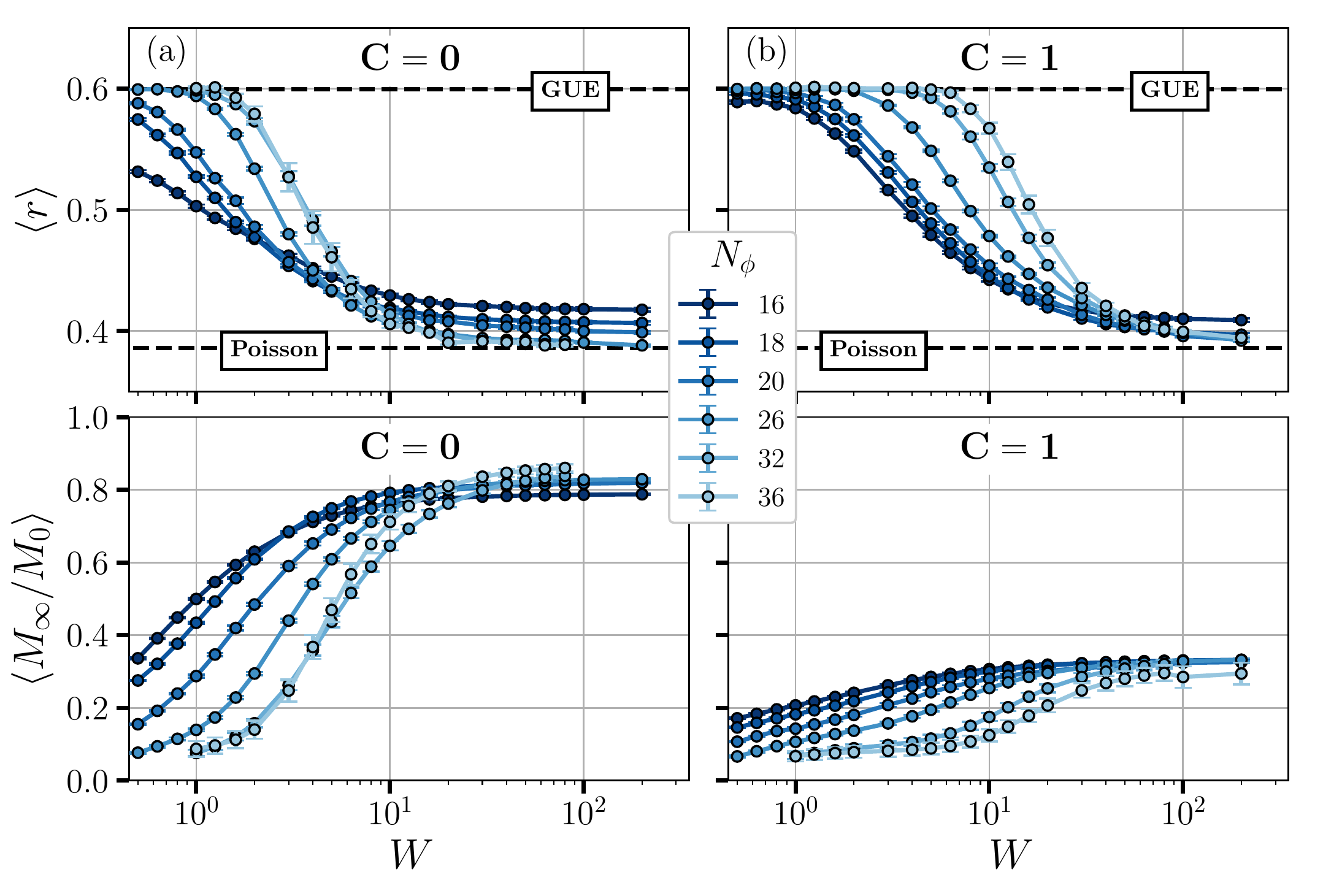}
    \end{minipage}
    \begin{minipage}{0.38\columnwidth}
    \includegraphics[ width=0.99\textwidth]{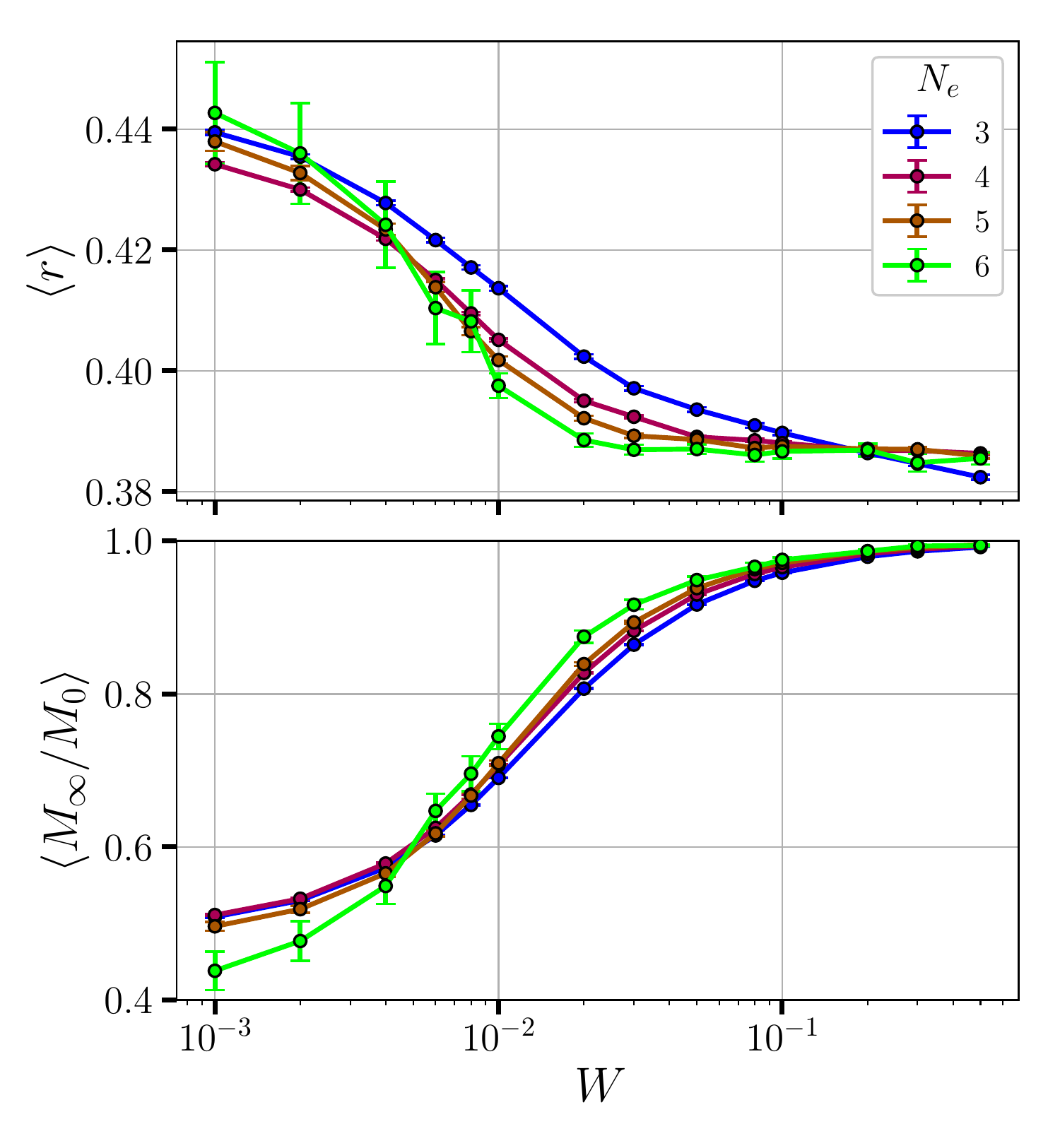}
    \end{minipage}
\caption{The $ r $ statistic (above) and remnant charge imbalance $M_\infty /  M_0 $ (below) are plotted against disorder for three different cases at a filling of $\nu = 1/3$.
In the left and middle columns, we have the $C=0$ and $C=1$ subbands of the periodic potential model. In the right column, we have the split-off states of the impurity band model.
The $ r $ statistic of the $C=0$ subband attains the localized Poisson value at much smaller disorder than the $C=1$ subband. The value of $ M_\infty / M_0 $ is also much larger for the $C=0$ subband, and also appears to flow towards a step function as system size is increased.
The $r$ statistic of the $C=1$ subband seems to grow rapidly with system size, similar the behavior seen in Fig.\ \ref{fig:Ger1}.
The remnant charge imbalance is small and reduces with system size, suggesting that the system has very little memory of initial conditions.
The $r$ statistic of the impurity band model is Poisson-like at large disorder, as expected, and seems to have a crossing at small disorder.
The remnant charge density of this model also crosses over at approximately the same critical disorder strength, and is nearly 1 in the large-disorder limit, suggesting nearly perfect retention of memory.
%In (c), we plot the critical disorder $W_c$ as a function of system size $N_\phi$.
%We define $W_c$ to be the value of W at which $ r  = 0.5$ (roughly halfway between Poisson and GUE) in the upper panel.
%In the lower panel, we define $W_c$ to be the value of W at which $\langle M_\infty / M_0 \rangle$ attains roughly half its saturation value.
%For the $C=0$ subband, we define $\langle M_\infty / M_0 \rangle (W_c) = 0.4$, and for the $C=1$ subband, we define $\langle M_\infty / M_0 \rangle (W_c) = 0.2$.
[Reprinted figure with permission from A.\ Krishna, M.\ Ippoliti, and R.\ N.\ Bhatt, Localization and interactions in topological and nontopological bands in two dimensions, Phys.\ Rev.\ B \textbf{100}, 054202 (2019).  Copyright (2019) by the American Physical Society.]
\label{fig:r_p2_q1_2d}}
\end{figure}

Note that the $M_0$ and $M_\infty$ denote expectation values over all possible outcomes for a \emph{single} realization of the disorder.
The ratio $M_\infty / M_0 \in [0, 1]$, called the `remnant charge imbalance', quantifies the extent to which the initial charge density modulation is `remembered' at infinite time.
In tandem with the $r$ statistic, this provides us with a second, useful metric to complement the level statistics to diagnose the lack of ergodicity and thus the possibility of a many-body localization transition.
For our analysis here, we ensemble average $M_\infty / M_0$ over all realizations of disorder at a value of $W$.

Fig.\ \ref{fig:r_p2_q1_2d} (left two panels) plots both these metrics $r$ and $M_\infty/ M_0$ as a function of disorder strength $W$ for the periodic potential model. Wherever possible, the curves are obtained at a filling $\nu = 1/3$. For values of $N_\phi$ that are not  multiples of 3, the curves are estimates obtained by interpolating between the nearest available rational fractions.
In the $C = 0$ subband, it is evident that there is a signature
of a finite disorder transition. At small $W$, the remnant charge
imbalance is close to zero, indicating a thermal phase in which
memory of initial conditions is washed away completely.
At large $W$, the remnant charge imbalance is nonzero. The
transition between the two regimes becomes sharper as the
system size is increased, indicating that the phenomenon is
likely to persist in the thermodynamic limit. This behavior
is mirrored in the eigenvalue $r$ statistic, which smoothly
interpolates between the GUE value at small disorder and the
Poisson value at large disorder with a crossing very close to
the Poisson value, in line with previous works \cite{Luitz2015}.

However, the topological $C = 1$ subband behaves in a starkly dissimilar manner. The remnant charge imbalance remains small, even at large values of disorder. As the system size is increased, it tends to become even smaller. 
The eigenvalue statistic $r$, while interpolating between its GUE and Poisson values, does so at much larger values of disorder. Importantly, it has no significant crossing as a function of system size, and the scaling of critical disorder $W_c$ with system size is just as in Fig.\ \ref{fig:Ger1} for the full LLL, suggesting no MBL in the thermodynamic limit.

For the impurity band model (right panel of Fig.\ \ref{fig:r_p2_q1_2d}), $\langle r \rangle$ statistic decreases with disorder $W$, as expected, and also shows a crossing (for the larger sizes) between different sizes around $W \approx 10^{-2}$.
The $\langle r \rangle$ value does not attain its GUE value of 0.6 as $ W \to 0$, because of the residual positional disorder of scatterers (governed by the $\rho$ parameter).
The remnant charge imbalance $\langle M_\infty / M_0 \rangle$ also shows a clear monotonically increasing trend as a function of disorder.
We observe a crossing in the curves at $W \approx 10^{-2}$, consistent with the eigenvalue statistics.
These suggest that there possibly is a finite disorder MBL transition in the non-topological split-off states in this system, just as for the $C=0$ subband in the periodic potential model.

All these studies, taken together, suggest that the behavior of a disordered interacting quantum Hall system is fundamentally different when the Hamiltonian is projected to a nontopological subspace than when projected to a topological subspace (or not projected at all).
The lack of access to larger system sizes precludes us from making a conclusive statement about whether this behavior is characteristic of a finite disorder transition for non-topological Landau level subbands in the thermodynamic limit.
It is possible that there is a slow drift of the crossing with system size, indicating the instability of true MBL in this system in the two-dimensional limit.
Further studies are necessary to clearly disambiguate the two scenarios.

We remark that rare fluctuation effects, which have been identified as destroying MBL in the thermodynamic limit for $d > 1$ are probably not playing a major role at our small sizes.
So this distinction between topological and non-topological bands has a different origin.

The focus of our results here is on two-dimensional scaling. 
In an earlier work \cite{Krishna2019}, we examined the behavior of similar models of Landau level subbands under quasi-dimensional scaling.
By keeping one side of the torus fixed and changing the length of the other side, we mimicked the effect of confining electrons to $d=1$, where MBL is more favored.
We found a marked dissimilarity in the behavior of topological and nontopological subbands in that case as well, reiterating that topology is of utmost importance in determining a system's propensity to many-body localize.

\section{Conclusions}

In the one and a half decades that have passed since the original paper by Basko et al.\ \citep{Basko2006} establishing the existence of many-body localization, starting from a completely Anderson localized band and including electron-electron interactions in a perturbative manner, it has become clear that MBL is much harder to achieve in interacting electron systems than Anderson localization in noninteracting models. The latter is ubiquitous in one dimension, and also in two dimensions with potential scattering, for any amount of disorder. The corresponding result at high magnetic fields yields localized states almost everywhere in the disorder-broadened Landau levels, except at critical energies (at least one per Landau level) where the localization length diverges.

In contrast, MBL is established conclusively only in one dimension for disordered spin chains \citep{Imbrie2016}. In fact, fairly strong arguments exist \citep{DeRoeck2017} suggesting that rare fluctuation effects destroy MBL is two dimensions and above. In this work, we have established that MBL does not exist in the thermodynamic limit for finite disorder for Landau levels in two dimensions, based on the evolution of the behavior of eigenvalue spacings with size even with rather modest sizes. 

In addition, by splitting the Landau level into topological and nontopological subbands using two vastly different protocols, and further reducing rare fluctuation effects so they do not play a major role at the sizes studied, we show that there is a marked contrast between the behavior of the two types of subbands with regard to many body localization.  This demonstrates that this contrasting behavior is primarily due to topology: nontopological bands do appear to be many-body localizable (in the absence of rare fluctuation effects) in the quantum Hall regime, whereas topological bands robustly resist MBL. This contrasting behavior is also seen in the long-time dynamics starting from a nonequilibrium state: while equilibrium is reached in the topological band, the non-topological band retains memory of the initial state till late times, another sign of MBL.

The question of what happens to MBL in the presence of long-ranged interactions is an open one.
Some arguments suggest that MBL does not exist in the presence of interactions longer ranged than $1/r^{2d}$ in $d$ dimensions (thus excluding the possibility of Coulomb interaction in any dimension) \cite{Yao2014}, while others suggest that MBL may arise even in 3-D systems with Coulomb interactions \cite{Nandkishore2017}.
We have not systematically investigated this issue in the context of our models.
All our results reported here considered a pure Haldane $V_1$ pseudopotential, which is as short-ranged an interaction as possible in the LLL. 
This choice was made due to our desire to provide the most conducive conditions for MBL to be stable, so that any variation in localization behavior could be attributed to the effects of topology alone.
However, we did perform a few numerical experiments using a Coulomb potential, and found that it led a much larger critical disorder strength $W_c$ for both topological and non-topological subbands.
This is in line with previous numerical work \cite{Tikhonov2018} indicating that long-ranged interactions make MBL harder to achieve.

\section{Acknowledgments}

Parts of the research described in this work were done in collaboration with Scott D.\ Geraedts and Matteo Ippoliti. Their contributions, and those of Rahul Nandkishore, to our overall understanding of many body localization in the quantum Hall regime are gratefully acknowledged.
In addition, RNB acknowledges the profound influence of the late P.\ W.\ Anderson on his research, particularly on disordered electronic systems, for over four decades.
This research was supported by the US Department of Energy, Division of Basic Energy Sciences through grant DE-SC0002140.

\bibliography{cas-refs}

\end{document}